\documentclass[structabstract]{aa}  
\usepackage{graphicx}
\usepackage{txfonts}
\usepackage{wrapfig, subfig, threeparttable, rotating}
\usepackage{natbib, lscape}   
\bibpunct{(}{)}{,}{a}{}{,} 
\begin{document}

   \title{Water distribution in shocked regions of the NGC1333-IRAS4A protostellar outflow
   }

   \author{G. Santangelo
          \inst{1,2}
          \and
          B. Nisini
          \inst{2}
          \and
          C. Codella
          \inst{1}
          \and
          A. Lorenzani
          \inst{1}
          \and
          U.~A. Yildiz
          \inst{3}
          \and
          S. Antoniucci
          \inst{2}
          \and
          P. Bjerkeli
          \inst{4,5,6}
          \and
          S. Cabrit
          \inst{7}
          \and
          T. Giannini
          \inst{2}
          \and
          L. Kristensen
          \inst{8}
          \and
          R. Liseau
          \inst{6}
          \and
          J. C. Mottram
          \inst{9}
          \and
          M. Tafalla
          \inst{10}
          \and
          E.~F. van Dishoeck
          \inst{9,11}
          }

   \institute{Osservatorio Astrofisico di Arcetri, Largo Enrico Fermi 5, 
              I-50125 Florence, Italy \\
              \email{gina@arcetri.astro.it}
         \and
              Osservatorio Astronomico di Roma, via di Frascati 33, 00040 Monteporzio Catone, Italy
         \and
              Jet Propulsion Laboratory, California Institute of Technology, 4800 Oak Grove Drive, Pasadena, CA, 91109, USA
         \and         
              Niels Bohr Institute, University of Copenhagen, Juliane Maries Vej 30, DK-2100 Copenhagen {\O}., Denmark
         \and
              Centre for Star and Planet Formation and Natural History Museum of Denmark, University of Copenhagen, {\O}ster Voldgade 5--7, DK-1350 Copenhagen K., Denmark
         \and
              Department of Earth and Space Sciences, Chalmers University of Technology, Onsala Space Observatory, 439 92 Onsala, Sweden
         \and
              LERMA, Observatoire de Paris, UMR 8112 of the CNRS, 61 Av. de l'Observatoire, 75014 Paris, France
         \and
              Harvard-Smithsonian Center for Astrophysics, 60 Garden Street, Cambridge, MA 02138, USA
         \and
              Leiden Observatory, Leiden University, P.O. Box 9513, 2300 RA Leiden, the Netherlands
         \and
              Observatorio Astron\'omico Nacional (IGN), Alfonso XII 3, E-28014 Madrid, Spain
         \and
              Max Planck Institut f{\"u}r Extraterrestrische Physik (MPE), Giessenbachstr.1, D-85748 Garching, Germany
             }

   \date{Received April 18, 2014; accepted June 20, 2014}

 
  \abstract
   {Water is a key molecule in protostellar environments because
its line emission is very sensitive
to both the chemistry and the physical conditions of the gas.
Observations of H$_2$O line emission from low-mass protostars 
and their associated outflows performed with HIFI onboard the 
\emph{Herschel} Space Observatory have highlighted
the complexity of H$_2$O line profiles, in which different kinematic components 
can be distinguished. 
}
   {The goal is to study the spatial distribution of H$_2$O, in particular 
of the different kinematic components detected in H$_2$O emission, at two bright shocked 
regions along IRAS4A, one of the strongest H$_2$O emitters among the Class 0 outflows.
}
   {We obtained \emph{Herschel}-PACS maps of the IRAS4A outflow and 
HIFI observations of two shocked positions. 
The largest HIFI beam of 38$^{\prime\prime}$ at 557~GHz was mapped in several key water lines 
with different upper energy levels, to reveal possible spatial variations of the line profiles. 
A large velocity gradient (LVG) analysis was performed to determine the excitation conditions 
of the gas.
}
   {We detect four H$_2$O lines and CO (16$-$15) at the two selected shocked positions. 
In addition, transitions from related outflow and envelope tracers are detected.
Different gas components associated with the shock are identified in the H$_2$O emission. 
In particular, at the head of the red lobe of the outflow, 
two distinct gas components with different excitation conditions are 
distinguished in the HIFI emission maps:
a compact component, detected in the ground-state water lines, and a more extended one.
Assuming that these two components correspond to two different temperature components observed 
in previous H$_2$O and CO studies, the LVG analysis of the H$_2$O emission suggests 
that the compact (about 3$^{\prime\prime}$, corresponding to about 700~AU) 
component is associated with a hot ($T$$\sim$1000~K) gas with densities 
$n_{\rm H_2}$$\sim$(1$-$4)$\times$$10^5$~cm$^{-3}$, 
whereas the extended ($10^{\prime\prime}$$-$$17^{\prime\prime}$, corresponding to 2400$-$4000~AU) 
one traces a warm ($T$$\sim$300$-$500~K) and dense gas
($n_{\rm H_2}$$\sim$(3$-$5)$\times$$10^7$~cm$^{-3}$).
Finally, using the CO (16$-$15) emission observed at R2 and assuming a typical CO/H$_2$ abundance of 
10$^{-4}$, we estimate the H$_2$O/H$_2$ abundance of the warm and hot components to be 
(7$-$10)$\times$10$^{-7}$ and (3$-$7)$\times$10$^{-5}$.
}
   {Our data allowed us, for the first time, to resolve spatially
the two temperature components previously observed with HIFI and PACS.
We propose that the compact hot component may be associated with the jet that impacts the surrounding material, 
whereas the warm, dense, and extended component originates 
from the compression of the ambient gas by the propagating flow.
}

   \keywords{Stars: formation -- Stars: low-mass -- ISM: jets and outflows -- ISM: individual objects: NGC1333-IRAS4A -- ISM: molecules
               }

   \titlerunning{Water distribution in shocked regions of NGC1333-IRAS4A}

   \maketitle
%

\section{Introduction}

Molecular outflows represent direct evidence of the earliest phases of star formation when collimated jets 
are driven \citep[e.g.,][]{arce2007,ray2007}. 
Shock fronts are generated at the point of impact of the ejected material with the surrounding cloud, 
inducing changes in the chemical composition and enhancing the abundance of several species.
Along with H$_2$ and CO, H$_2$O is one of the main cooling agents in shocks 
\citep[][]{kaufman1996,flower2010,karska2013}.
It is also very sensitive to physical conditions and chemical processes of the shocked gas 
\citep[][]{vandishoeck2011}.
An increase of the water gas-phase abundance from $<$10$^{-7}$ up to 3$\times$10$^{-4}$ is expected in warm shocked gas 
($\gtrsim$100~K) because of the combined effects of sputtering of ice mantles 
and high-temperature reactions that drive atomic oxygen into H$_2$O 
\citep[][]{hollenbach1989,kaufman1996,flower2010,suutarinen2014}.

Systematic observations of multiple H$_2$O transitions in prototypical star-forming regions
have been conducted with \emph{Herschel} \citep{pilbratt2010}.
Thanks to its much higher sensitivity and spectral resolution and its smaller beam size 
with respect to previous space missions, such as ISO, \emph{Odin}, and SWAS, 
\emph{Herschel} has provided strong constraints on the water abundance and 
physical conditions in water-emitting gas.
In particular, the Water In Star-forming regions with \emph{Herschel} 
\citep[WISH,][]{vandishoeck2011}
key program has been dedicated to the study of the physical and dynamical properties of water and 
its role in shock chemistry.
The H$_2$O line profiles observed with the Heterodyne Instrument for the Far Infrared 
\citep[HIFI,][]{degraauw2010} 
at outflow shocks show several kinematic components along with 
variations with excitation energy \citep{kristensen2012,vasta2012,santangelo2012}. 
The observed H$_2$O emission probes warm ($>$200~K) and very dense gas 
($n_{\rm H_2}$$\gtrsim$$10^6$~cm$^{-3}$), which is associated with high-$J$ CO emission 
\citep[e.g][]{karska2013,santangelo2013}
and is not traced by other molecules seen from the ground, such as low-$J$ CO and SiO 
\citep[e.g.,][]{vasta2012,nisini2013,tafalla2013,santangelo2013}. 
The differences in line profiles of the various tracers also confirm
the uniqueness of H$_2$O as a probe of shocked gas. 

A two-temperature-components model has been suggested by \citet{santangelo2013} 
to reproduce the H$_2$O, CO, and mid-IR H$_2$ lines, observed 
along the L1448 and L1157 protostellar outflows. 
This model consists of 
1) an extended warm component ($T$$\sim$500~K) traced by lower excitation H$_2$O emission 
($E_{\rm u}$$\lesssim$140~K) 
and by CO lines up to $J$$=$22$-$21; and 2) a compact hot component ($T$$\sim$1000~K) 
traced by higher excitation H$_2$O emission and higher-$J$ CO transitions. 
Two gas components with different excitation conditions have also been 
proposed by \citet{busquet2014} to account for the H$_2$O and CO emission observed at the 
bright shock region B1 in the L1157 outflow.
Finally, multiple-temperature 
components have been suggested to explain 
the spectrally unresolved CO ladder
at the position of several Class 0 sources 
\citep[e.g.,][]{goicoechea2012,herczeg2012,dionatos2013,karska2013,manoj2013,green2013}.
The nature of these two components and, notably, their spatial extent 
has not yet been clarified, however.
To understand this problem better, maps of velocity-resolved H$_2$O lines that 
are sensitive to different excitation conditions are needed.

NGC1333 \citep[$d$=235~pc,][]{hirota2008} is a well-studied star-forming region and contains 
many young stellar objects (YSOs) and outflows 
\citep[e.g.,][]{liseau1988,bally1996,kneesandell2000}.
Within the region, IRAS4A and IRAS4B are two low-mass protostars 
with a projected separation of about 30$^{\prime\prime}$; 
both have been identified as binary systems using mm interferometry 
\citep[e.g.,][]{lay1995,looney2000,choi2005,jorgensen2007}.
The companion to IRAS4B is detected at a separation of 11$^{\prime\prime}$, 
whereas IRAS4A is resolved into two components with a separation of only 2$^{\prime\prime}$. 
IRAS4A is one of the youngest protobinary systems found so far, 
as inferred by 
its strong dust continuum emission 
with cold blackbody-like spectral energy distribution and its well-collimated
outflow \citep[e.g.,][]{sandell1991,blake1995}, extending over arcminute scales.
The IRAS4A low-mass protostar has been the subject of extensive 
observations with ground-based submillimeter telescopes and interferometers
\citep[e.g.,][]{blake1995,difrancesco2001,maret2005,choi2005,jorgensen2007,yildiz2012}.
As part of WISH, \citet{kristensen2010,kristensen2012} observed several H$_2$O transitions 
towards the IRAS4A source with \emph{Herschel}-HIFI, showing the complex line profiles 
with multiple components within the HIFI beam. In the line profiles, these authors identified 
a broad Gaussian component (FWHM$\gtrsim$20~km~s$^{-1}$) that
was also detected in the CO (10$-$9) emission 
\citep{yildiz2013} and is
associated with the molecular outflow. In addition, they identified a so-called medium-broad component, 
offset with respect to the source velocity and with smaller line widths (FWHM$\sim$5$-$10~km~s$^{-1}$), 
which they associated with currently shocked gas close to the
protostar \citep{kristensen2013}.

In this paper, we present new \emph{Herschel}-PACS and HIFI observations 
of several key H$_2$O lines that are sensitive to different excitation conditions, 
and HIFI CO (16$-$15) spectra at two shocked positions along the IRAS4A outflow. 
The data are complemented by ground-based CO (3$-$2) and (6$-$5) maps by \citet{yildiz2012}.
The goal is to study the spatial distribution of the water emission to spatially separate the 
multiple kinematic components that were previously detected towards the source within the HIFI beam.
The observations and data reduction are described in Sect.~\ref{sect:observations}. 
In Sect.~\ref{sect:results} we present the observational results. The analysis and interpretation 
of the H$_2$O excitation conditions and its physical origin are discussed in 
Sect.~\ref{sect:water_components}.
Finally, in Sect.~\ref{sect:conclusions}, we present the main conclusions.


\section{Observations}
\label{sect:observations}

\subsection{PACS observations}
\label{subsect:PACS_obs}

\begin{figure*}
\centering
\includegraphics[width=0.85\textwidth]{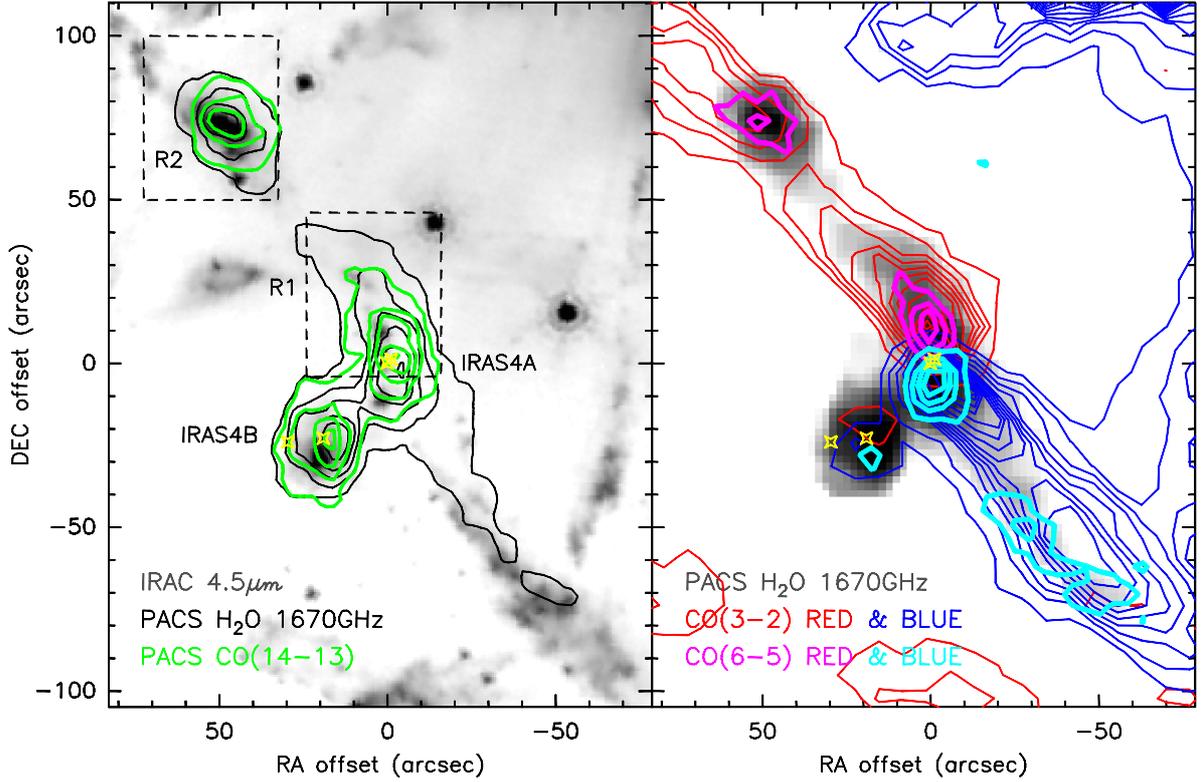}
\caption{PACS map of the H$_2$O ($2_{12}$$-$$1_{01}$) emission at 1670~GHz of the IRAS4 region
compared with the Spitzer-IRAC emission at 4.5~$\mu$m and 
the CO (14$-$13) emission in the left panel and with the 
JCMT CO (3$-$2) and APEX CO (6$-$5) emission \citep{yildiz2012} in the right panel.
The CO (3$-$2) and CO (6$-$5) maps are integrated in the velocity ranges between 
-20~km~s$^{-1}$ and 3~km~s$^{-1}$ 
for the blue-shifted emission and 12 km~s$^{-1}$ and 50~km~s$^{-1}$ for the red-shifted emission.
The contour levels start at the 5~$\sigma$ level and increase in steps of 10~$\sigma$ for 
the PACS H$_2$O and CO (14$-$13) maps, from the 5~$\sigma$ level emission in steps of 5~$\sigma$ 
for the CO (3$-$2) and from the 5~$\sigma$ level emission in steps of 3~$\sigma$ for the CO (6$-$5).
Offsets are with respect to the central source IRAS4A, at coordinates 
$\alpha_{J2000}$$=$03$^{\rm h}$29$^{\rm m}$10$.\!\!^{s}$50, 
$\delta_{J2000}$$=$+31$^\circ$13$^{\prime}$30$.\!\!^{\prime\prime}$9.
The positions of the IRAS4A and IRAS4B binary sources are marked with 
yellow symbols \citep{looney2000}. 
The regions mapped with HIFI in the water lines around the selected shock positions (R1 and R2) 
are indicated.
}
\label{fig:maps}
\end{figure*}

PACS maps of H$_2$O ($2_{12}$$-$$1_{01}$) at 1670~GHz and CO (14$-$13) emission 
were used for the analysis. The maps are presented in Fig.~\ref{fig:maps},
and a summary of the observations is given in Table~\ref{table:PACS_lines}. 
The observations are part of the OT1 program
\emph{``Probing the physics and dynamics of the hidden warm gas in the youngest protostellar
outflows''} (OT1\_bnisini\_1).
The PACS instrument is an Integral Field Unit (IFU), consisting of a 5$\times$5 
array of spatial pixels, each covering 9$.\!\!^{\prime\prime}$4$\times$9$.\!\!^{\prime\prime}$4, 
for a total field of view of 47$^{\prime\prime}$$\times$47$^{\prime\prime}$.
PACS was used in line-spectroscopy 
mode to obtain a spectral Nyquist-sampled raster map of IRAS4A. The reference coordinates are 
at the position of IRAS4A,
$\alpha_{J2000}$$=$03$^{\rm h}$29$^{\rm m}$10$.\!\!^{s}$50, 
$\delta_{J2000}$$=$+31$^\circ$13$^{\prime}$30$.\!\!^{\prime\prime}$9.
The diffraction-limited FWHM beam size at 179~$\mu$m is about 13$^{\prime\prime}$.
The data were reduced with {\sc hipe}\footnote{{\sc hipe} is a joint development 
by the \emph{Herschel} Science Ground Segment Consortium, consisting of ESA, the NASA \emph{Herschel} 
Science Center, and the HIFI, PACS and SPIRE consortia.} 
\citep[\emph{Herschel} Interactive Processing Environment,][]{ott2010} version~9.0. 
Within {\sc hipe}, they were flat-fielded and flux-calibrated
by comparison with observations of Neptune. The calibration uncertainty is estimated to be 
around 20\%, based on the flux repeatability for multiple observations of the same target 
in different programs and on cross-calibration with HIFI and ISO.
Finally, continuum subtraction was performed in {\sc IDL}, and integrated line maps were obtained.

\begin{table}
\caption{Parameters of the lines mapped with PACS.} 
\label{table:PACS_lines}
\centering
\begin{threeparttable}
\renewcommand{\footnoterule}{}
\begin{tabular}{l c c c c } 
\hline\hline
Line & Frequency & Wavelength & $E_{\rm u}/k_{\rm B}$ & HPBW \\ 
& (GHz) & ($\mu$m) & (K) & (arcsec) \\ 
\hline
o-H$_2$O ($2_{12}$$-$$1_{01}$) & 1669.90 & 179.5 & 114.4 & 13  \\ 
CO (14$-$13)             & 1611.79 & 186.0 & 580.5 & 13  \\ 
\hline
\end{tabular}
\end{threeparttable}
\tablefoot{
The observation IDs of both lines are 1342225852, 1342225853, and 1342225854.
}
\end{table}

\subsection{HIFI observations}
\label{subsect:HIFI_obs}

\begin{table*}
\caption{Parameters of the lines observed with HIFI.}
\label{table:HIFI_lines}
\centering
\begin{threeparttable}
\renewcommand{\footnoterule}{}
\begin{tabular}{l c c c c c r r c } 
\hline\hline
Line & Observation ID & Band & Mode\tnote{a} & Sideband & $\eta_{\rm MB}$ & Frequency & $E_{\rm u}/k_{\rm B}$ & HPBW \\
& &&& && (MHz) & (K) & (arcsec) \\ 
\hline
o-H$_2$O ($1_{10}$$-$$1_{01}$) & 1342248895, 1342248897 & 1 & SP & LSB & 0.75 &  556936.07   &  61 & 38  \\ 
p-H$_2$O ($2_{11}$$-$$2_{02}$) & 1342249432, 1342249433 & 2 &  M & USB & 0.75 &  752033.23   &  137 & 28  \\
o-H$_2$O ($3_{12}$$-$$3_{03}$) & 1342249853, 1342249854 & 4 &  M & USB & 0.74 & 1097365.05   &  249 & 19   \\
p-H$_2$O ($1_{11}$$-$$0_{00}$) & 1342249021, 1342250209 & 4 &  M & USB & 0.74 & 1113343.06   & 53 & 19   \\ 
CO (16$-$15)             & 1342249639, 1342249640 & 7 & SP & USB & 0.70 & 1841345.51   &      752 & 12   \\ 
\hline
\end{tabular}
\begin{tablenotes}
\item[a] SP = Single Pointing mode; M = Mapping mode. 
\end{tablenotes}
\end{threeparttable}
\end{table*}

\begin{figure*}
\centering
\includegraphics[width=0.9\textwidth]{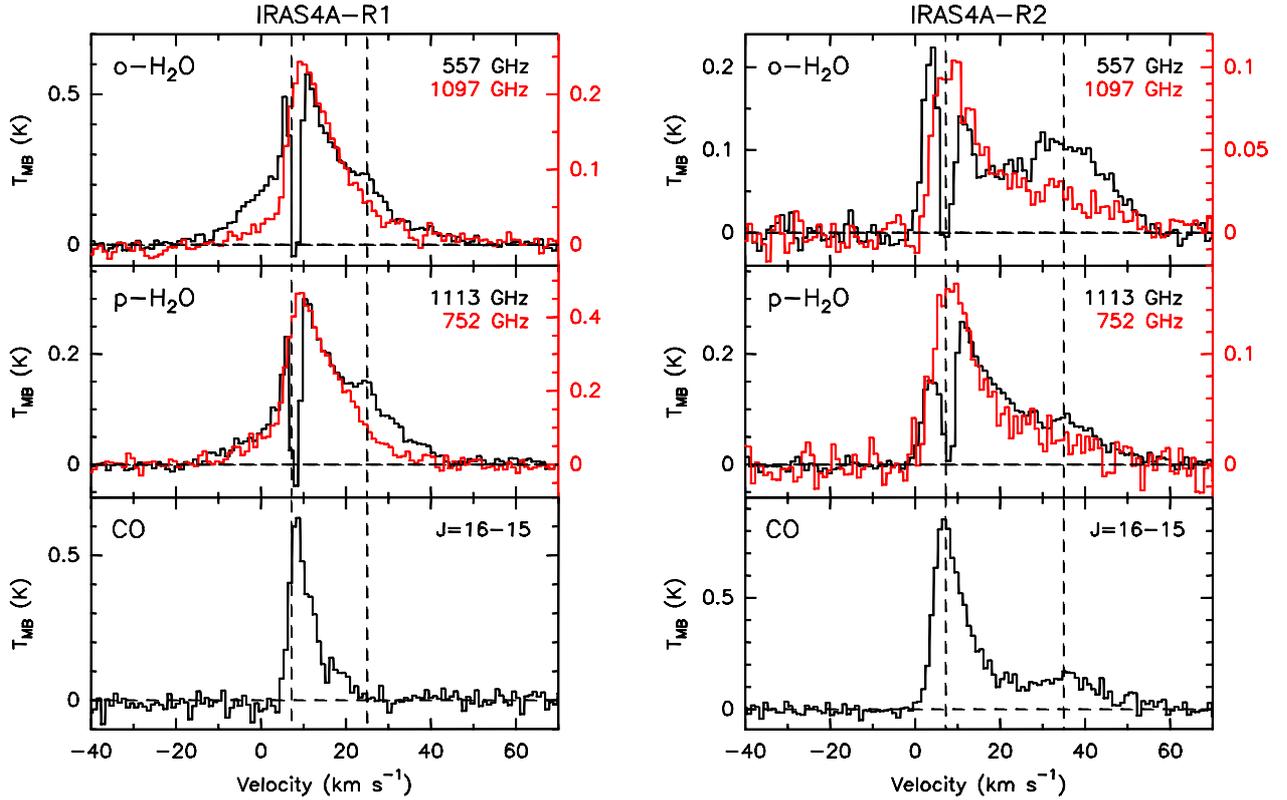}
\caption{HIFI spectra of the H$_2$O and CO transitions observed 
at the R1 (left) and R2 (right) shock positions along IRAS4A.
The spectra are all convolved to the same angular resolution of the H$_2$O ($1_{10}$$-$$1_{01}$) line at 557~GHz 
($\sim$$38^{\prime\prime}$), with the exception of the 
CO (16$-$15) line, which is a single pointing HIFI observation at $\sim$$12^{\prime\prime}$ 
resolution.
The spectra shown in red have intensities provided in the right-hand axes.
The vertical dashed line marks the systemic velocity ($v_{\rm LSR}$$=$+7.2~km~s$^{-1}$) and the H$_2$O 
secondary emission peaks at +25 and +35~km~s$^{-1}$ for R1 and R2.
}
\label{fig:spettri}
\end{figure*}

\begin{figure}
\centering
\includegraphics[angle=-90,width=0.4\textwidth]{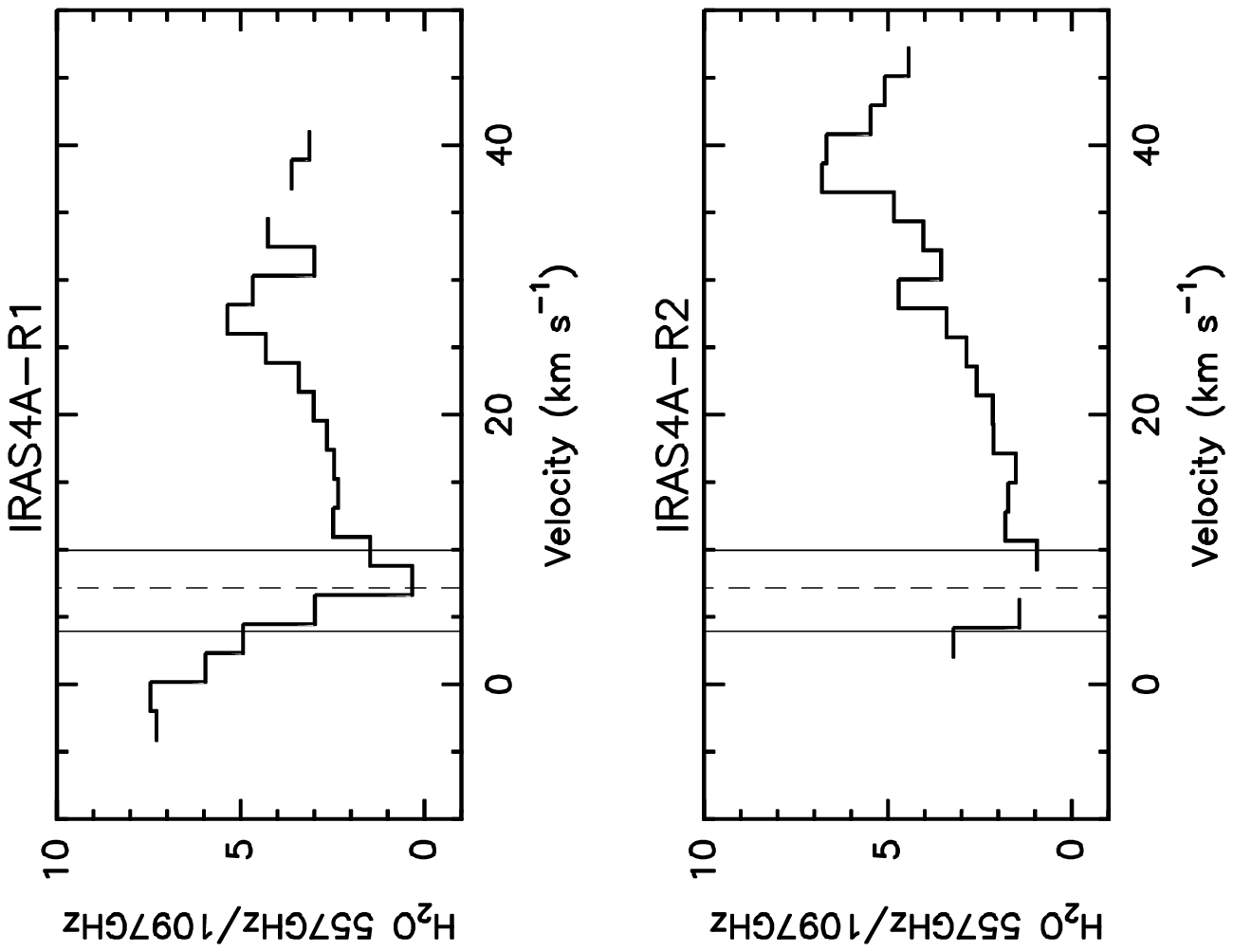}
\caption{Ratio between the o-H$_2$O ($1_{10}$$-$$1_{01}$) (at 557~GHz) and 
o-H$_2$O ($3_{12}$$-$$3_{03}$) (at 1097~GHz) 
lines as a function of velocity at the two observed shock positions R1 (upper panel) and R2 (lower panel). 
The 1097~GHz water spectra are convolved to the same angular resolution as the 557~GHz spectra, 
i.e., 38$^{\prime\prime}$.
The line ratios are plotted only where the $S/N$ ratio is higher than three for the two lines, 
at a spectral resolution of 2~km~s$^{-1}$. The vertical dashed line indicates the source velocity, 
whereas the solid vertical lines mark the velocity range 
of the absorption dip (4$-$10~km~s$^{-1}$).
}
\label{fig:H2Olineratios}
\end{figure}

We selected two active shock positions along the IRAS4A outflow: 
R1 ($\alpha_{J2000}$$=$03$^{\rm h}$29$^{\rm m}$10$.\!\!^{s}$82, 
$\delta_{J2000}$$=$+31$^\circ$13$^{\prime}$51$.\!\!^{\prime\prime}$9), 
at the origin of the jet from the driving source, and R2 
($\alpha_{J2000}$$=$03$^{\rm h}$29$^{\rm m}$14$.\!\!^{s}$59, 
$\delta_{J2000}$$=$+31$^\circ$14$^{\prime}$45$.\!\!^{\prime\prime}$8), 
which is the head of the red lobe of the outflow.
They appear as very bright peaks in the PACS H$_2$O ($2_{12}$$-$$1_{01}$) and CO (14$-$13) maps 
shown in Fig.~\ref{fig:maps}.

Single-pointing observations of the o-H$_2$O ($1_{10}$$-$$1_{01}$) line at 557~GHz and the CO (16$-$15)
transition at the two selected positions were conducted with \emph{Herschel}-HIFI
in dual beam-switch and fast-chop mode.
In addition, 
an area of size equal to the HIFI beam width at 557~GHz (38$^{\prime\prime}$)
was mapped in on-the-fly mode in three other H$_2$O lines,
spanning excitation energies $E_{\rm u}$ from 50~K to 250~K (see Table~\ref{table:HIFI_lines}).
The observations were carried out between July 2012 and August 2012 
as part of the OT2 program
\emph{``Solving the puzzle of water excitation in shocks''} (OT2\_gsantang\_1).
Contextually, the spectral set-up allowed us to observe transitions from other molecules: 
N$_2$H$^+$ (6$-$5), SO ($13_{14}$$-$$12_{13}$), CH$_3$OH ($3_{-2}$$-$$2_{-1}$), NH$_3$ (1$_0$$-$$0_0$), and 
$^{13}$CO (10$-$9).
A summary of the performed observations is given in Tables~\ref{table:HIFI_lines} 
and \ref{table:HIFI_additional_lines}. 

The data were processed with the ESA-supported package {\sc hipe} version 11 for calibration.
The calibration uncertainty is taken to be 20\%.
Further reduction of all the spectra, including baseline subtraction, and 
the analysis of the data were 
performed using the GILDAS\footnote{http://www.iram.fr/IRAMFR/GILDAS/} software.
H- and V-polarizations were co-added after inspection to increase sensitivity, 
since no significant differences were found 
between the two data sets.
The calibrated $T^*_A$ scale from the telescope was converted into the $T_{\rm mb}$ scale using the 
main-beam efficiency factors provided by \citet{roelfsema2012}\footnote{See also
http://herschel.esac.esa.int/twiki/bin/view/Public/\\
HifiCalibrationWeb?template=viewprint}
and reported in Table~\ref{table:HIFI_lines}.
At the velocity resolution of 1~km~s$^{-1}$, the rms noise ranges between 10~mK and 20~mK ($T_{\rm mb}$ scale).

\begin{figure*}
\centering
\includegraphics[width=0.43\textwidth]{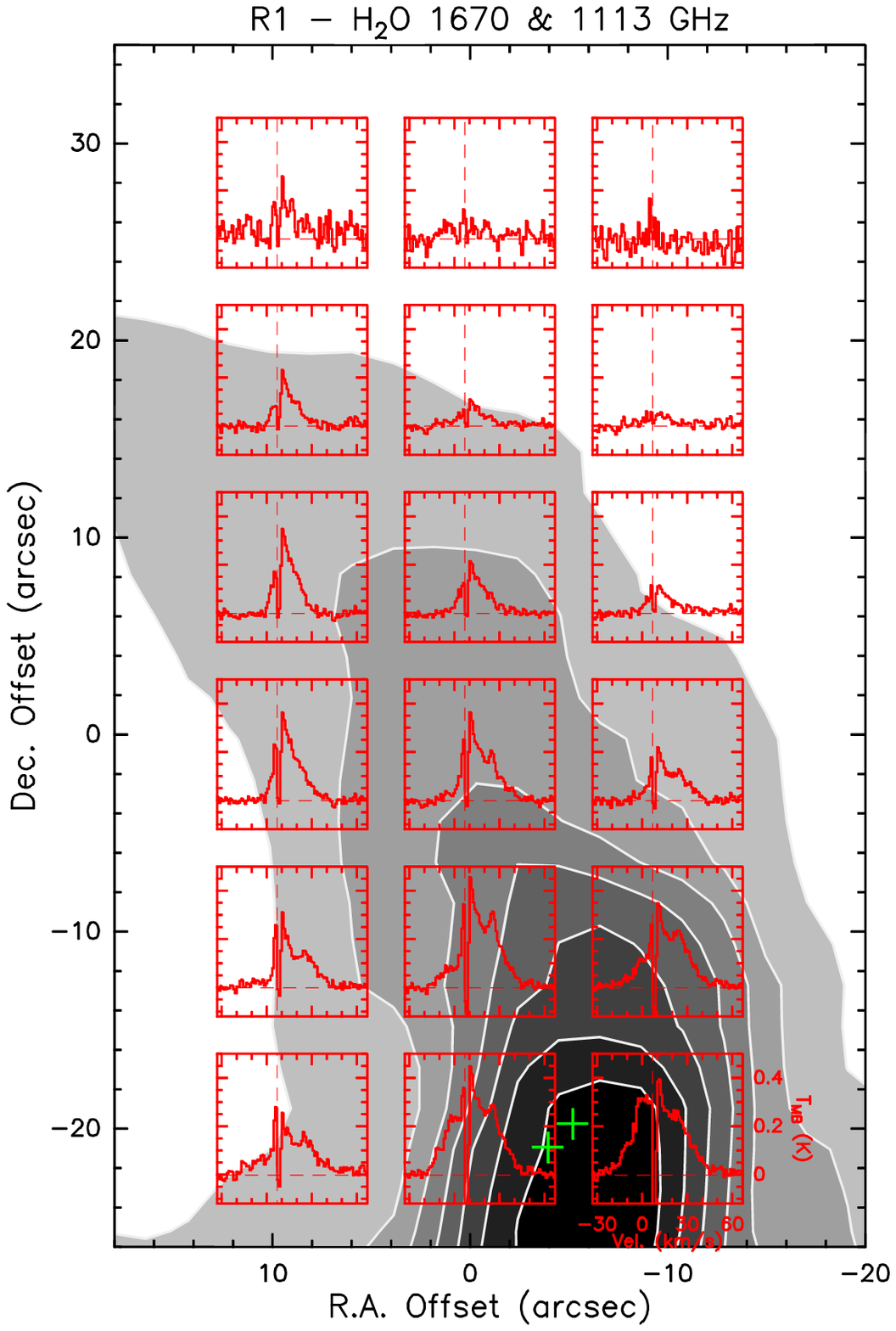}
\includegraphics[width=0.43\textwidth]{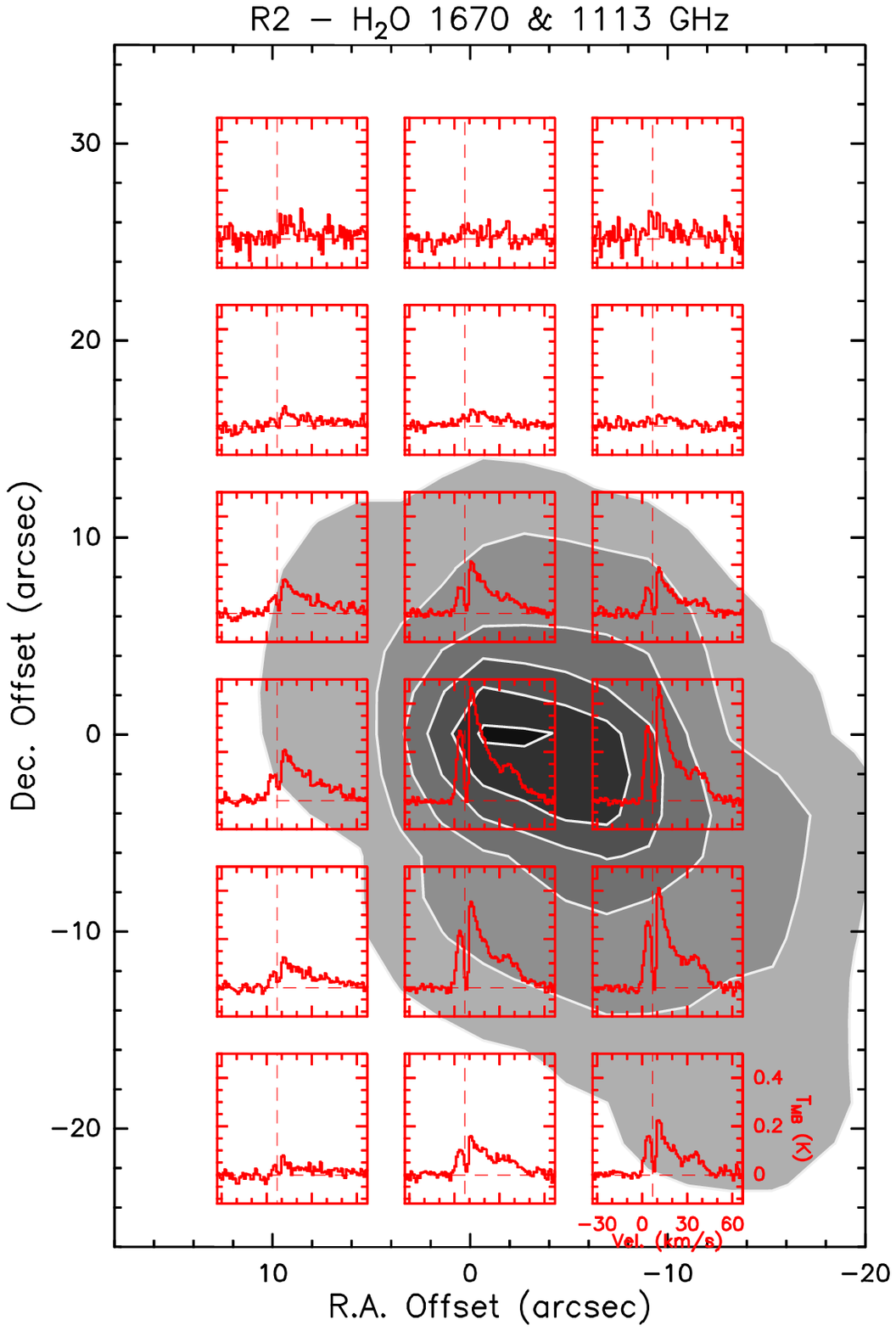}
\caption{HIFI spectra of the 1113~GHz H$_2$O line, mapped at R1 (left) and R2 (right). 
The spectra are overlaid
on the respective PACS H$_2$O map at 1670~GHz (grey scale and white contours). 
Offsets are with respect to the R1 and R2 shock positions.
The IRAS4A binary source position is marked with green crosses.
}
\label{fig:mapEspettri}
\end{figure*}


\section{Results}
\label{sect:results}

Figure~\ref{fig:maps} shows the PACS maps of H$_2$O ($2_{12}$$-$$1_{01}$) at 1670~GHz 
and CO (14$-$13), in comparison with the Spitzer-IRAC emission at 4.5~$\mu$m 
and the ground-based CO (3$-$2) and CO (6$-$5) \citep{yildiz2012} towards the IRAS4 region.
The figure highlights the spatial correlation between H$_2$O, high-$J$ CO, and Spitzer 4.5~$\mu$m emissions, 
in agreement with previous studies of outflows from low-mass protostars 
\citep[e.g.,][]{nisini2010,santangelo2013,tafalla2013}. 
In particular, PACS CO (14$-$13) and APEX CO (6$-$5) emissions are spatially associated with 
the PACS H$_2$O emission, whereas the low-$J$ CO (3$-$2) emission is more extended and offset.
A sharp change of propagation direction of $\sim$$30^{\circ}$ in the north-eastern outflow red lobe, 
occurring close to the R1 shock position, is visible both in H$_2$O and CO emission.
This directional variability was previously shown by both single-dish and interferometric observations,
and several mechanisms were proposed to explain it, 
including magnetic deflection, a precessing jet, and collisions with a dense core in the ambient cloud
\citep[e.g.,][]{blake1995,girart1999,choi2001,choi2005,baek2009,choi2011,yildiz2012}.
Strong H$_2$O emission peaks are found at the location of active shocked regions 
and at the position of IRAS4A and its neighbour IRAS4B \citep[see also][]{nisini2010,nisini2013}.
In particular, the R1 and R2 shock positions appear as bright peaks in the H$_2$O 
and high-$J$ CO emission, as revealed by PACS and ground-based observations.

An overview of the HIFI observations is given in Fig.~\ref{fig:spettri} and Fig.~\ref{afig:OTHERlines}, 
where H$_2$O and CO spectra observed at the two selected shock positions 
and the spectra of additional lines detected with HIFI are shown. 
All H$_2$O spectra observed in mapping mode are convolved to the same angular resolution 
of 38$^{\prime\prime}$ for comparison with the observations of the ground-state o-H$_2$O transition at 557~GHz.
Several kinematic components
can be distinguished in the observed H$_2$O line profiles at both the R1 and R2 positions. 
First, an absorption dip around the systemic velocity of IRAS4A 
\citep[$v_{\rm LSR}$$=$+7.2~km~s$^{-1}$,][]{kristensen2012} is detected in the ground-state transitions 
of o- and p-H$_2$O (at 557 GHz and 1113 GHz), associated with cold gas in the outer envelope. 
Second, a triangularly shaped outflow wing is present up to about 50~km~s$^{-1}$ at R1 and about 60~km~s$^{-1}$ at R2. 
Third, an excess of emission at high velocity is observed in the ground-state transitions of o- and p-H$_2$O.
This secondary high-velocity (HV)
emission peak appears at a velocity of about +25~km~s$^{-1}$ and +35~km~s$^{-1}$ in R1 and R2.
Similar variations of water line profiles with excitation were observed at the 
bow-shock positions along the red lobes of the L1448 (R4) and L1157 (R) outflows
by \citet{santangelo2012} and \citet{vasta2012}.

Figure~\ref{fig:H2Olineratios} presents 
the H$_2$O 557~GHz/1097~GHz line ratio as a function of velocity for R1 and R2.
Water transitions with different upper level energies were chosen and convolved 
to the same angular resolution of 38$^{\prime\prime}$, and the ratios are plotted 
only for velocities where both transitions have $S/N$$>$3.
Neglecting the velocity range 4$-$10~km~s$^{-1}$, where the absorption dip 
in the 557~GHz H$_2$O contaminates the analysis, the ratio between the o-H$_2$O ($1_{10}$$-$$1_{01}$) 
and o-H$_2$O ($3_{12}$$-$$3_{03}$) lines increases significantly with velocity, reflecting 
the fact that the secondary HV peak at both R1 and R2 appears in the lower excitation 
energy transitions (see Fig.~\ref{fig:spettri}). Finally, we note that the 
increasing trend is stronger at R2, the shock position farthest from the central driving source.
We point out that these findings for the shock positions are in contrast to 
observations of H$_2$O emission lines at the central protostellar positions, which show constant line ratios 
(Mottram et al., in prep.).

Recent studies have shown that high-$J$ CO emission is associated with H$_2$O emission, corresponding to a warm 
($\gtrsim$300~K) and dense ($n_{\rm H_2}$$\gtrsim$$10^6$~cm$^{-3}$) gas component
\citep[e.g.,][]{karska2013,santangelo2013}.
This finding is consistent with our CO (16$-$15) observations at R2, showing 
a similar line profile as the ground-state water transitions, that is, strong, broad emission around the systemic 
velocity of the source and an HV emission peak.
We note that the detection of a secondary HV emission component at R2 in the H$_2$O 557~GHz 
and in the CO (16$-$15) lines, with angular resolutions of about 38$^{\prime\prime}$ and 12$^{\prime\prime}$, 
suggests that this emission is associated with a compact gas component centred on the shock.
Therefore, the non-detection of this HV
emission peak in the higher-excitation, smaller-beam size H$_2$O transitions is probably due to 
an excitation effect, with the HV component being less excited than the low-velocity (LV) component, 
as previously found in the L1448 and L1157 bow shock positions \citep[e.g.,][]{vasta2012,santangelo2012}.

On the other hand, at the R1 position 
no clear secondary peak around +25~km~s$^{-1}$ is detected in CO (16$-$15).
This difference possibly arises because the HV peak is spatially shifted towards IRAS4A by $\sim$10$^{\prime\prime}$ 
with respect to R1, while the CO (16$-$15) was observed with a beam size of 12$^{\prime\prime}$
(see also Sect.~\ref{subsect:water_distribution} and Figs.~\ref{fig:mappette_H2O} and \ref{fig:mappette_CO}).

\subsection{Water spatial distribution}
\label{subsect:water_distribution}

Figure~\ref{fig:mapEspettri} shows the HIFI spectra at 1113~GHz observed 
around the R1 and R2 shock positions. 
The figure highlights the variation of the water line profiles around the shocked regions.

\begin{figure}
\centering
\includegraphics[angle=-90,width=0.43\textwidth]{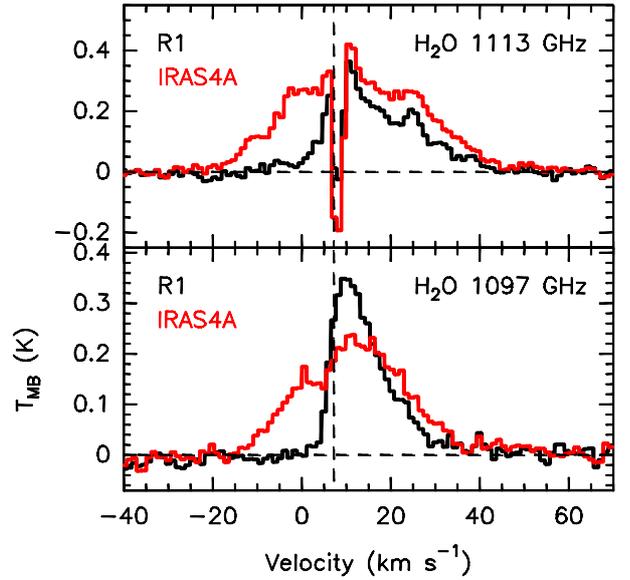} 
\caption{Comparison between the H$_2$O ($1_{11}$$-$$0_{00}$) (1113~GHz) and ($3_{12}$$-$$3_{03}$) (1097~GHz) 
line profiles at R1 and on the IRAS4A source. The vertical dashed line indicates the source velocity.
}
\label{fig:water_iras4a_r1}
\end{figure}

From examining the R1 position, we note that the H$_2$O line profiles close to IRAS4A resemble 
those along the outflow, which testifies that the outflow dominates the water profiles.
This similarity is even clearer when we compare the H$_2$O profiles centred at R1 and at IRAS4A 
(Fig.~\ref{fig:water_iras4a_r1}).
The p-H$_2$O 1113~GHz spectra show the same profile in the red-shifted emission, although 
the on-source position is brighter than R1.
On the other hand, the H$_2$O profiles of the higher excitation energy H$_2$O 1097~GHz line 
appear to be different, with the on-source position being fainter than R1.
This difference suggests that at R1 the excitation conditions of the H$_2$O emitting gas 
are different from the conditions at the position of the central source.
In addition, in the on-source spectra, significant blue-shifted emission is detected that is
not seen at the R1 position.
The blue wing presents an excess of emission at $\sim$0~km~s$^{-1}$ 
($-7$~km~s$^{-1}$ with respect to the source velocity), 
which is quite symmetric in the 1113~GHz line with respect to the red-shifted secondary peak.
\citet{kristensen2013} interpreted this blue-shifted H$_2$O component as 
originating from a compact dissociative shock close to the protostar.
We remark that our profiles present some differences with those presented 
in \citet{kristensen2013}; in the latter, for example, the secondary red-shifted peak is not as bright 
and the relative intensity of the low- and high-velocity blue-shifted components 
is significantly different. 
These differences are probably due to the different observation pointings. 
However, \citet{kristensen2013} suggested that time variability might also change the H$_2$O line profiles.

\begin{figure*}
\centering
\includegraphics[angle=-90,width=0.48\textwidth]{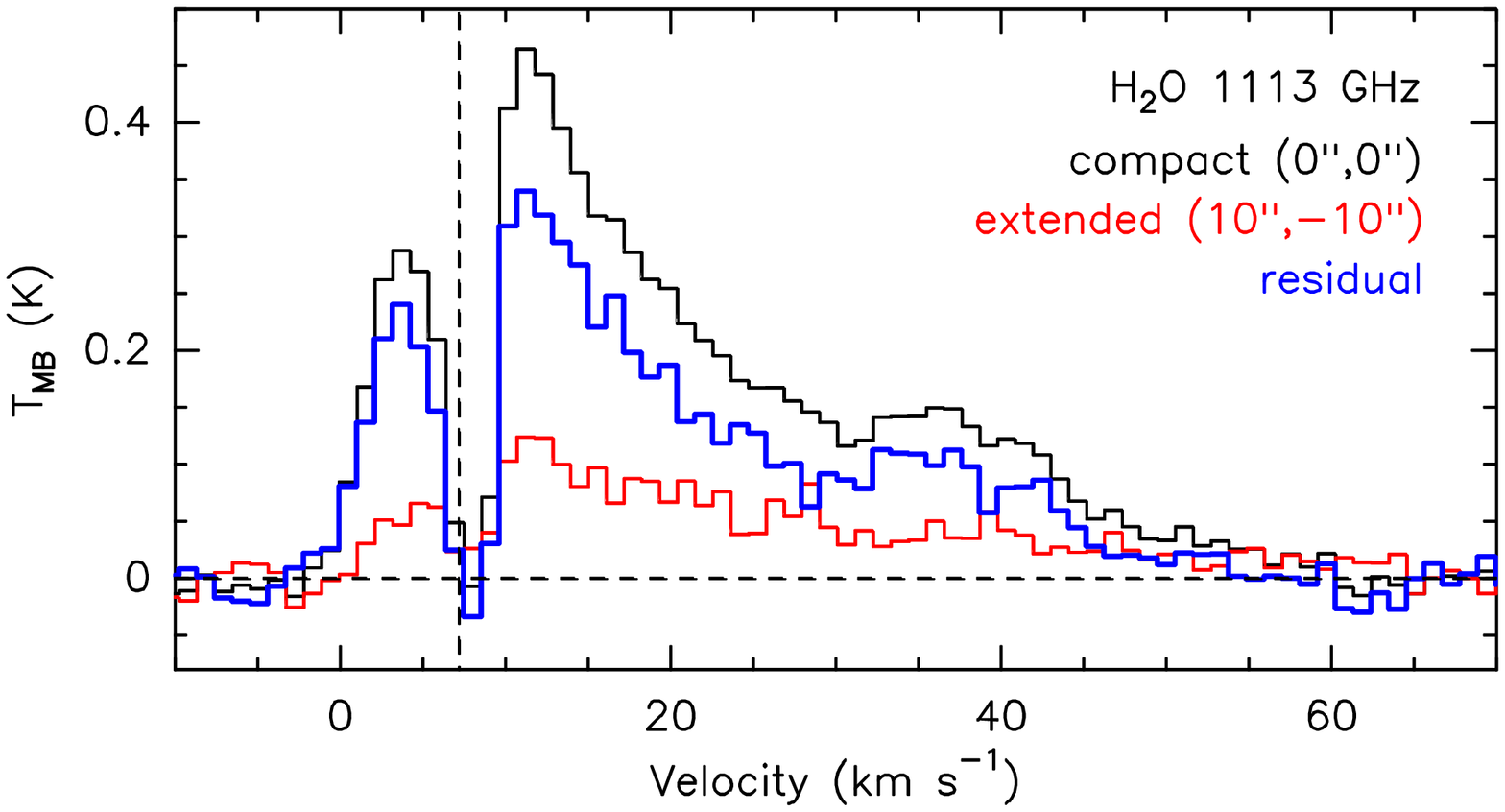}
\includegraphics[angle=-90,width=0.47\textwidth]{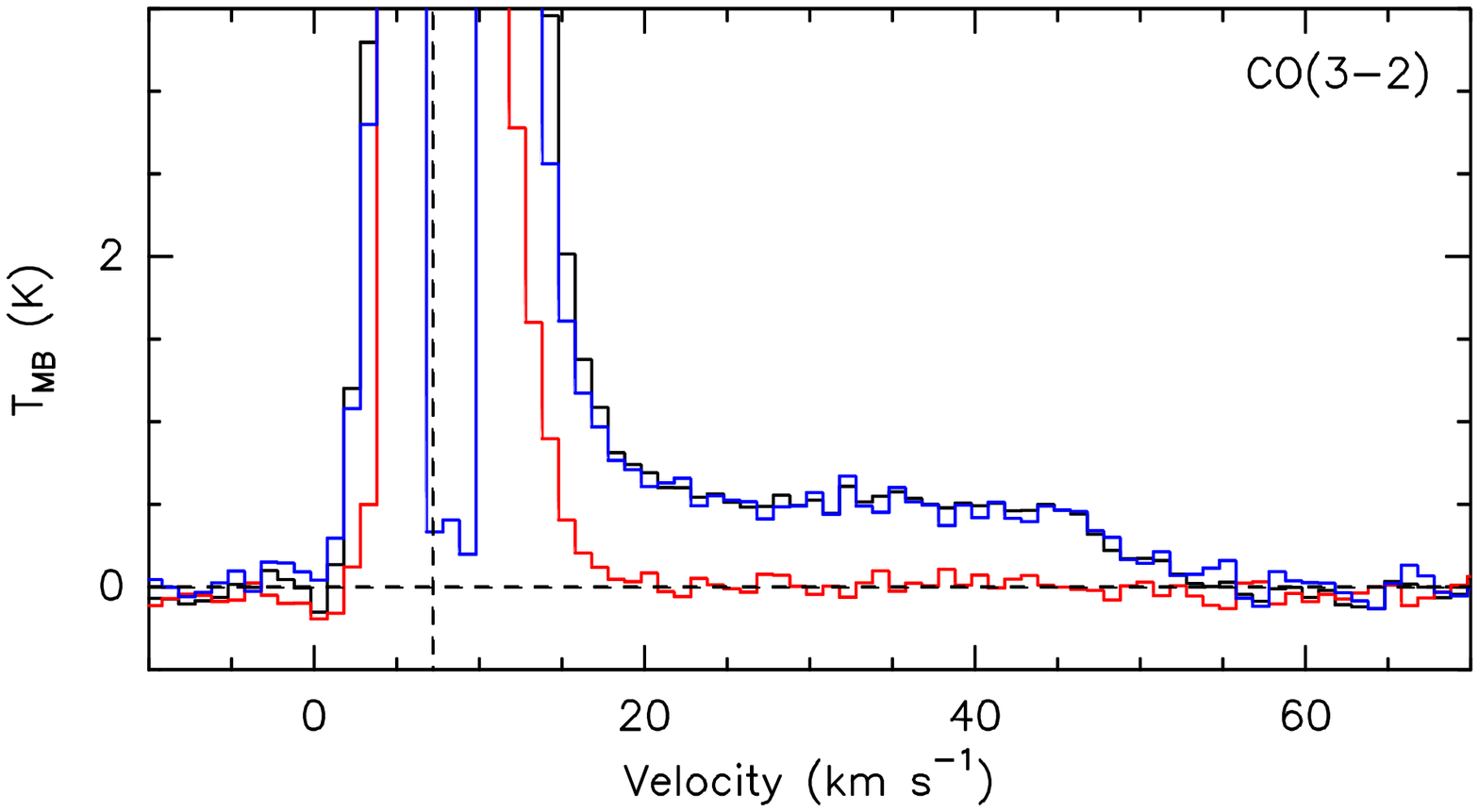}
\caption{\emph{Left}: comparison between the H$_2$O $1_{11}$$-$$0_{00}$ emission at 1113~GHz 
(19$^{\prime\prime}$ beam size) at the R2 shock position (black), 
corresponding to the compact gas component, and that at the offset position 
(10$^{\prime\prime}$, $-$10$^{\prime\prime}$)
with respect to R2 (red), corresponding to the extended component. 
The residual spectrum, 
given by the difference between the two displayed spectra, is shown in blue.
Offsets are with respect to R2.
The vertical dashed line indicates the source velocity.
\emph{Right}: same as the upper panel for the CO (3$-$2) emission.
}
\label{fig:r2_profiles}
\end{figure*}

The R2 position, which is a pure shock position far from the driving source of the outflow, 
is a very interesting laboratory to study the water distribution around shocks.
Our HIFI observations clearly show that 
the different water kinematic components mentioned 
above (see Sect.~\ref{sect:results} and Fig.~\ref{fig:spettri})
are not uniformly distributed across the mapped region (Fig.~\ref{fig:mapEspettri}).
In particular, the HV peak 
is observed only at the bright shock peak, but is not detected in 
water line profiles offset from the shock position, which show
a triangular wing shape. 
These two types of line profiles probably indicate two 
different gas components in the water emission: 
a compact gas component, located at the shock peak and associated with the HV peak, and a more diffuse component.
The difference is shown in the left panel of Fig.~\ref{fig:r2_profiles}, 
where line emission from the R2 position and a position ($10^{\prime\prime}$,$-$$10^{\prime\prime}$)
offset from the R2 shock are chosen as representative. An emission residual between 
the two spectra is also displayed.
The spectrum associated with the compact component 
shows an excess of emission at the systemic velocity
and at the HV peak with respect to the extended component.
The two detected components are therefore not kinematically distinct, since 
the velocity range of the emission is the same in both cases. 
Thus, it would not have been possible to distinguish them within the 38$^{\prime\prime}$ 
beam size of single-pointing observations. 

\begin{figure*}
\centering
\includegraphics[width=0.8\textwidth]{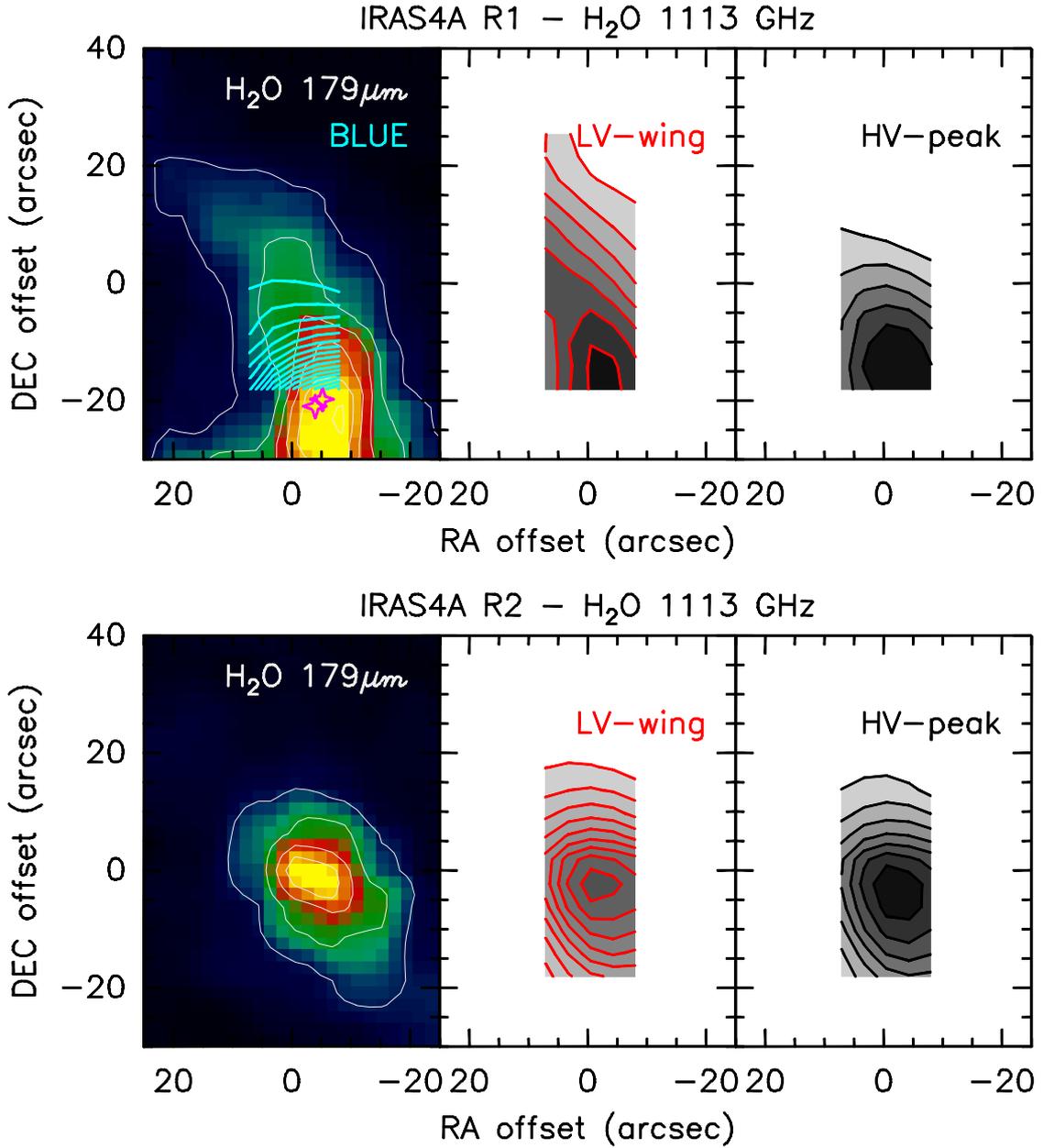}
\caption{Velocity-integrated maps of the HIFI H$_2$O 1113~GHz emission at the R1 (upper panel)
and R2 (lower panel) positions compared with the PACS H$_2$O 179~$\mu$m map.
At R1, the H$_2$O 1113~GHz emission is integrated in three velocity ranges: 
the blue-shifted emission (between $-20$~km~s$^{-1}$ and 3~km~s$^{-1}$), the low-velocity (LV) wing emission 
(between 11~km~s$^{-1}$ and 20~km~s$^{-1}$), and the high-velocity (HV) emission peak (between 20~km~s$^{-1}$ 
and 30~km~s$^{-1}$). At R2, two velocity ranges are considered: the LV wing emission 
(between 11~km~s$^{-1}$ and 30~km~s$^{-1}$) and the HV emission peak (between 30~km~s$^{-1}$ and 45~km~s$^{-1}$).
The contour levels start from the 5~$\sigma$ level and increase in steps of 10~$\sigma$ for the PACS 
179~$\mu$m emission and in steps of 3~$\sigma$ for the HIFI 1113~GHz emission.
Offsets are with respect to the R1 and R2 shock positions.
The magenta symbols represent the position of the IRAS4A binary source.
}
\label{fig:mappette_H2O}
\end{figure*}

In Fig.~\ref{fig:mappette_H2O}, the maps of the H$_2$O 1113~GHz emission, 
integrated in the three velocity ranges reported in the caption and 
corresponding to blue-shifted emission, LV red-wing emission, 
and HV emission peak, are presented for the R1 (upper panel) and the R2 (lower panel) positions, 
in comparison with the H$_2$O 1670~GHz PACS map.
Blue-shifted emission is only detected close to the central source IRAS4A. 
At R1, the HV-peak emission appears to have a compact distribution, spatially 
associated with IRAS4A, whereas the LV-wing emission is elongated in the outflow red-lobe direction.
In contrast, no significant difference between the LV-wing and the 
HV-peak emission can be observed at R2 at this angular resolution (19$^{\prime\prime}$, corresponding to about 4500~AU) 
with both distributions appearing unresolved. 
This difference is discussed in more detail in Sect.~\ref{sect:water_components}.

\begin{figure*}
\centering
\includegraphics[width=0.8\textwidth]{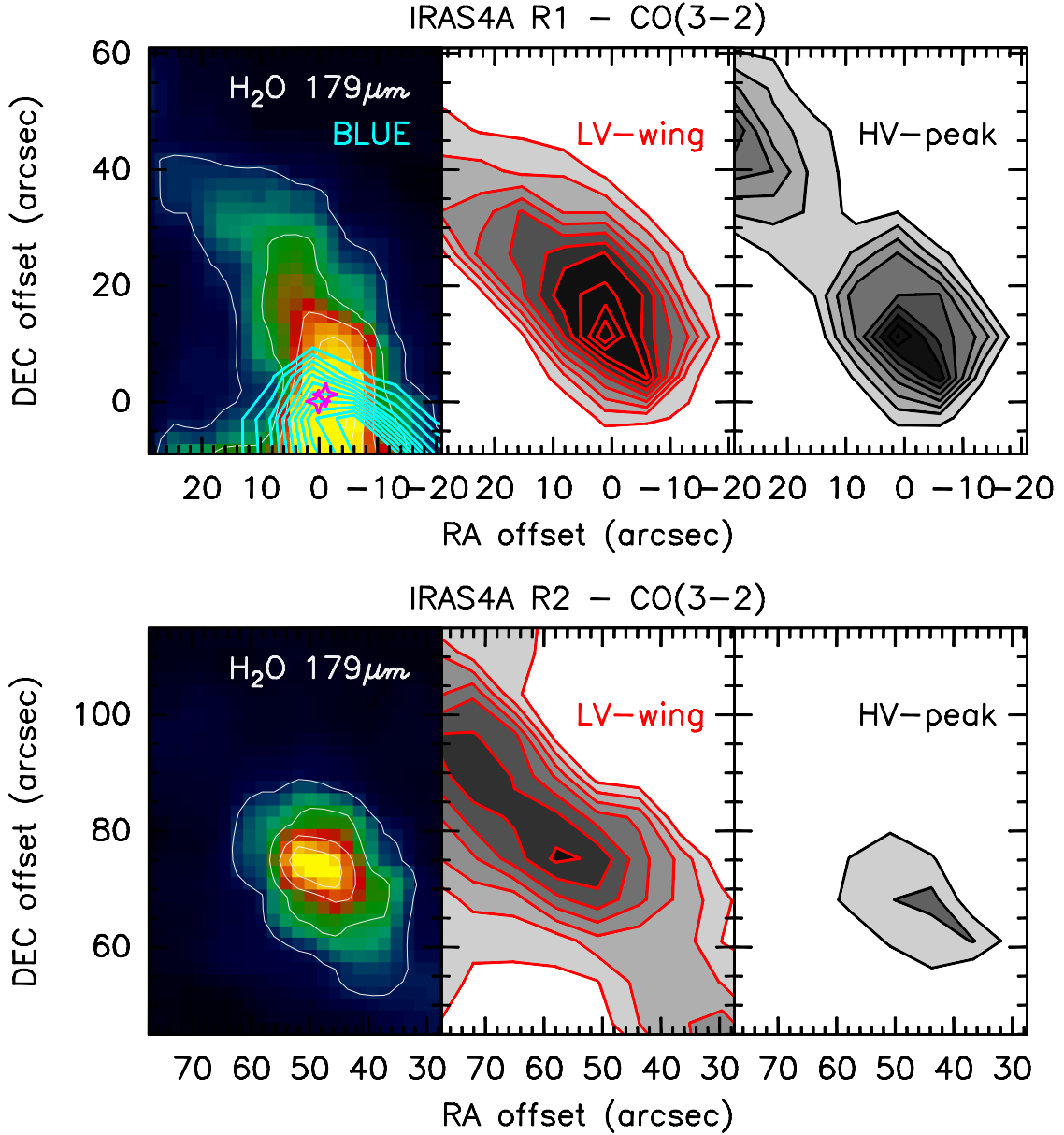}
\caption{Same as Fig.~\ref{fig:mappette_H2O} for the CO (3$-$2) emission.
Offsets here are with respect to the central source IRAS4A.
}
\label{fig:mappette_CO}
\end{figure*}

Similar conclusions can be drawn at R1 from Fig.~\ref{fig:mappette_CO}, where 
the CO (3$-$2) emission is integrated in the same velocity ranges 
as adopted for the H$_2$O maps of Fig.~\ref{fig:mappette_H2O}.
The HV-peak emission appears to be more compact than the LV-wing emission and shifted with respect to R1 towards IRAS4A.
We note that a second north-eastern emission peak can be identified in the HV-peak emission, 
which is not associated with bright PACS H$_2$O 179~$\mu$m emission.
A different situation with respect to the H$_2$O 1113~GHz emission
is observed at R2 in the CO (3$-$2) emission, however. Here, the HV-peak emission 
shows a compact distribution that is not resolved on an angular scale of 14$^{\prime\prime}$ 
and spatially associated with the H$_2$O emission at 
179~$\mu$m. In contrast, the LV-wing emission is associated with a diffuse gas component,
following the outflow direction, and is spatially more extended than the H$_2$O emission. 
The CO (3$-$2) observations thus seem to support the scenario 
of two distinct gas components in the R2 shock position. 

Finally, a comparison between the CO (3$-$2) spectra observed at R2 and 
at a position (10$^{\prime\prime}$, $-$10$^{\prime\prime}$) offset from the shock 
is presented in the right panel of Fig.~\ref{fig:r2_profiles}.
Although the CO (3$-$2) spectra show different line profiles in the LV emission 
with respect to H$_2$O, once more, they indicate that the HV peak is associated with a 
compact gas component not detected in the more diffuse LV gas.

We note that, because of the low sensitivity in the line wings, the 
APEX CO (6$-$5) data do not detect HV line emission and thus cannot be used to analyse the two spatial components 
observed in the H$_2$O and CO (3$-$2) lines.

\begin{figure}
\centering
\includegraphics[width=0.45\textwidth]{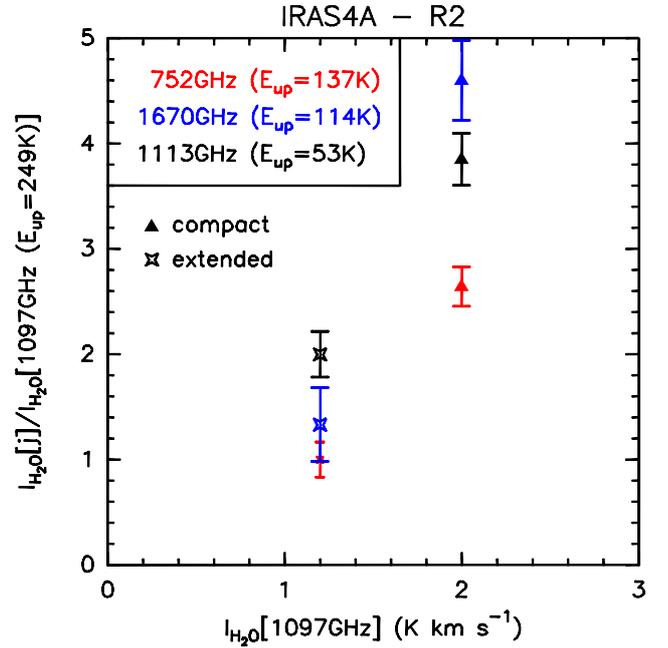}
\caption{H$_2$O line ratios with respect to the o-H$_2$O ($3_{12}$$-$$3_{03}$) line at 1097~GHz
measured at the R2 shock position, corresponding to the compact component, and 
at the offset position (10$^{\prime\prime}$, $-$10$^{\prime\prime}$) with respect to R2, corresponding 
to the extended component.
The HIFI water line intensities are integrated in the line wings (see Fig.~\ref{fig:LVG}) 
at the original angular resolution of the spectra 
(see Table~\ref{table:HIFI_lines}), 
while the intensity of the H$_2$O line at 1670~GHz in the two gas components 
is measured from the PACS map by averaging over 19$^{\prime\prime}$. 
The intensity of the H$_2$O ($2_{11}$$-$$2_{02}$) line at 752~GHz 
associated with the compact component is 
corrected for the beam size ratio with respect to the 1097~GHz line (28$^{\prime\prime}$/19$^{\prime\prime}$), 
assuming the source to be point-like.
}
\label{fig:r2_ratios}
\end{figure}


\section{Two gas components in shocked H$_2$O emission}
\label{sect:water_components}

To investigate the presence of two distinct gas components at R2, 
we compared their associated H$_2$O line ratios with respect to the higher excitation energy 
H$_2$O 1097~GHz line (Fig.~\ref{fig:r2_ratios}).
We assumed that the compact gas component 
dominates the H$_2$O emission at the R2 shock position 
while the extended component dominates the H$_2$O emission at the position 
(10$^{\prime\prime}$, $-$10$^{\prime\prime}$) offset from R2, as supported by the 
difference in the observed line profiles (see Figs.~\ref{fig:mapEspettri} and \ref{fig:r2_profiles}).
The line intensities are integrated only in the line wings, between 10~km~s$^{-1}$ and 60~km~s$^{-1}$, 
because of the absorption dip at the source systemic velocity. 
The two distinct gas components with different excitation conditions at R2 
are confirmed by the significantly different associated water line ratios. 
In particular, the three measured line ratios 
are higher in the compact component, which is consistent with this component 
being less excited than the extended one and detected only in the lower excitation water lines.

\begin{figure*}
\centering
\includegraphics[width=0.8\textwidth]{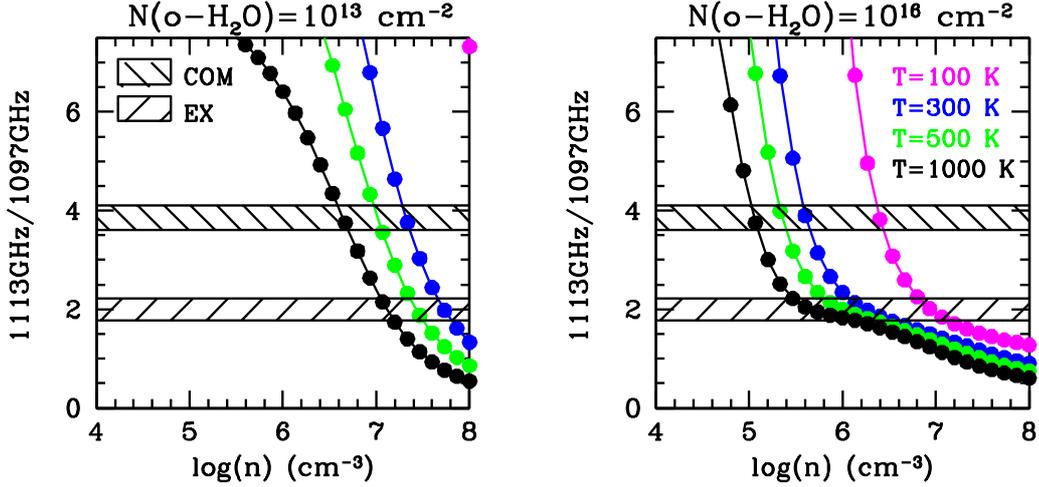}
\caption{{\sc radex} predictions for 
the ratio between the p-H$_2$O ($1_{11}$$-$$0_{00}$) (at 1113~GHz) and the o-H$_2$O ($3_{12}$$-$$3_{03}$) 
(at 1097~GHz) lines as a function of the H$_2$ density for four 
values of kinetic temperature (100, 300, 500, and 1000~K) and 
two values of o-H$_2$O column density (10$^{13}$ and 10$^{16}$~cm$^{-2}$).
The shaded bands represent the water ratios observed 
for the compact and extended components at R2, i.e., the R2 position and a position 
(10$^{\prime\prime}$, $-$10$^{\prime\prime}$) offset from R2.
The HIFI water line intensities are integrated 
in the line wings, between 10 and 60~km~s$^{-1}$, 
at the original angular resolution of the spectra (19$^{\prime\prime}$). 
}
\label{fig:LVG}
\end{figure*}

To characterize the two gas components at R2 in terms of excitation conditions, 
we ran the {\sc radex} non-LTE molecular LVG radiative transfer code \citep{vandertak2007} in 
plane-parallel geometry, with collisional rate coefficients from 
\citet{dubernet2006,dubernet2009} and \citet{daniel2010,daniel2011} and molecular data from 
the Leiden Atomic and Molecular Database 
\citep[LAMDA\footnote{http://www.strw.leidenuniv.nl/$\sim$moldata/},][]{schoier2005}. 
A grid of models with density ranging between 
10$^4$~cm$^{-3}$ and 10$^8$~cm$^{-3}$ and temperatures ranging between 100~K and 1600~K was built.
Two values of o-H$_2$O column density were considered, corresponding to optically thin and moderately thick water emission.
We adopted a typical line width of 20~km~s$^{-1}$ for both components from the HIFI spectra 
and an H$_2$O ortho-to-para ratio equal to 3 \citep{emprechtinger2013}, corresponding to the 
high-temperature equilibrium value. 
The ratio between the p-H$_2$O 1113~GHz and the o-H$_2$O 1097~GHz lines with 
similar beam sizes (19$^{\prime\prime}$) was considered to avoid beam-filling problems.
The line-wing intensities (between 10~km~s$^{-1}$ and 60~km~s$^{-1}$) were measured at the same chosen positions 
corresponding to the two gas components.
A comparison between the observed and predicted water line ratios as a function of the H$_2$ density 
for four values of temperature (100~K, 300~K, 500~K, and 1000~K) 
is presented in Fig.~\ref{fig:LVG}.
In both cases, the low-density regime can be excluded, meaning that a gas density $n$$\gtrsim$$10^5$~cm$^{-3}$
is required. 
Such high densities are consistent with HIFI CS (12$-$11) observations\footnote{The data are part of the 
OT1 program ``Peering into the protostellar shocks: NH$_3$ emission at high-velocities''.}
at a similar position about (9$^{\prime\prime}$, 4$^{\prime\prime}$) offset from R1 
(G{\'o}mez-Ruiz et al., in prep.), suggesting that
gas densities in excess of 10$^{5}$~cm$^{-3}$ are needed to reproduce the CS (12$-$11) intensity in the line wing. 
They are also consistent with JCMT observations of broad CS (10$-$9) emission 
at the position of the IRAS4A central source, probing warm and dense gas \citep{jorgensen2005}.

To investigate the excitation conditions of the two H$_2$O components at R2 in detail, 
we assumed that these components correspond to the warm and hot components 
found in previous works, that is, the warm 
component at about 300$-$500~K and the hot 
component at about 1000~K \citep[see][]{santangelo2013,busquet2014}. 
By assuming these temperatures and modelling the observed 
H$_2$O 1113/1097~GHz line ratio and the intensity of the H$_2$O 1097~GHz line 
using the {\sc radex} LVG code, 
we find that the hot component is associated with small emitting sizes (about 3$^{\prime\prime}$, 
corresponding to about 700~AU), 
gas densities $n_{\rm H_2}$$\sim$(1$-$4)$\times$$10^5$~cm$^{-3}$, and o-H$_2$O column densities 
$\sim$(0.5$-$1)$\times$$10^{16}$~cm$^{-2}$ (corresponding to 
$\tau_{\rm 557~GHz}$$\sim$15$-$40 at the peak of the H$_2$O emission). 
The warm component is instead more extended (sizes of about $10^{\prime\prime}$$-$$17^{\prime\prime}$, 
corresponding to 2400$-$4000~AU)
and associated with higher gas densities $n_{\rm H_2}$$\sim$(3$-$5)$\times$$10^7$~cm$^{-3}$ and lower 
o-H$_2$O column densities $\sim$(1$-$2)$\times$10$^{13}$~cm$^{-2}$ 
(corresponding to optically thin H$_2$O emission, with $\tau_{\rm 557~GHz}$$\sim$0.04).
The derived sizes for the extended and compact components 
are consistent with the fact that we cannot spatially resolve 
and separate them in the H$_2$O emission with angular resolutions higher than 19$^{\prime\prime}$
(see Fig.~\ref{fig:mappette_H2O}), while they can be spatially distinguished in the CO (3$-$2) 
emission, where the angular resolution is higher (14$^{\prime\prime}$, see Fig.~\ref{fig:mappette_CO}).

Next, we used the CO (16$-$15) emission observed at R2 to estimate 
the H$_2$O abundance of the two spatial components. 
Since similar line profiles are observed for CO (16$-$15) and H$_2$O at this position
(see Sect.~\ref{sect:results} and Fig.~\ref{fig:spettri}), we can assume that they share 
a common origin, meaning that the excitation conditions of CO (16$-$15) are the same 
as derived for H$_2$O in both components.
Since the CO (16$-$15) spectrum at R2 is a single-pointing observation, however, we cannot 
spatially separate the emission from the two components as we did with H$_2$O. 
Although the two components are not kinematically distinct 
(see Sect.~\ref{subsect:water_distribution}), 
a crude way of separating their relative contribution to the intensity of the 
single observed CO (16$-$15) spectrum and thus deriving rough estimates of the H$_2$O abundance, 
is to separate them in velocity; hence,
we attribute the velocity range from +11~km~s$^{-1}$ to +30~km~s$^{-1}$ to the extended 
component and from 30~km~s$^{-1}$ to 45~km~s$^{-1}$ to the compact one, 
as suggested by the velocity-integrated maps of CO (3$-$2) 
(Fig.~\ref{fig:mappette_CO}) and the CO (16$-$15) line profile.
Assuming for each component the same excitation conditions and emission sizes as derived from H$_2$O, 
we obtain CO column densities of about 2$\times$10$^{15}$~cm$^{-2}$ and 2$\times$10$^{16}$~cm$^{-2}$ 
for the extended and compact components, respectively, which correspond to an H$_2$O/H$_2$ abundance 
of about (7$-$10)$\times$10$^{-7}$ for the warm extended component 
and (3$-$7)$\times$10$^{-5}$ for the hot compact component 
(assuming a typical CO/H$_2$ abundance of 10$^{-4}$).
The low fractional H$_2$O abundance associated with the warm gas component 
agrees with other studies of molecular outflows
\citep[e.g.,][]{bjerkeli2012,vasta2012,nisini2013,santangelo2013,tafalla2013,busquet2014}.
Moreover, our finding of higher fractional H$_2$O abundance in the hot gas 
is consistent with ISO data \citep[e.g.,][]{giannini2001} and previous \emph{Herschel} observations 
of shocked gas along the L1448 and L1157 outflows \citep[e.g.,][]{santangelo2013,busquet2014}.

We speculate that the compact hot component, detected in the H$_2$O emission at R2, 
may be associated with the jet that impacts the surrounding material. 
Conversely, the warm, dense, and extended component originates 
from the compression of the ambient gas by the propagating flow.
This picture was recently proposed by \citet{busquet2014} for the L1157 outflow. 
Our data, however, allow for the first time to spatially resolve
these 
two gas components through emission maps and confirm this scenario.
We point out that high-angular resolution observations are crucial to probe the
structure of the investigated shock region in depth.

%

\section{Conclusions}
\label{sect:conclusions}

We performed \emph{Herschel}-HIFI observations of two shock positions (R1 and R2) along the IRAS4A outflow.
An area corresponding to the size of the largest HIFI beam of 38$^{\prime\prime}$ at 557~GHz 
was mapped in several key water lines with different upper level energies 
to study the water spatial distribution and to separate spatially 
different gas components associated with the shock. 
The main results of the work can be summarized as follows:
\begin{enumerate}
\item At both selected shock positions, we detect four H$_2$O lines with upper energy levels 
in the range 50$-$250~K and CO (16$-$15).
In addition, transitions from related outflow and envelope tracers are detected.
\item At the R2 shock position, the head of the red-lobe of the 
outflow, two gas components with different excitation conditions are detected from the HIFI maps: 
a compact component, detected in the ground-state water lines, and a more extended component.
They are not kinematically distinct, since the velocity range 
of the emission is similar in both cases, thus they cannot be distinguished within the 
large \emph{Herschel} beam sizes of 19$^{\prime\prime}$ and 38$^{\prime\prime}$ 
at the frequencies of the ground-state H$_2$O transitions.
\item The LVG analysis of the H$_2$O emission suggests 
that the compact (about 3$^{\prime\prime}$) component is associated with a hot ($T$$\sim$1000~K) 
gas with densities $n_{\rm H_2}$$\sim$(1$-$4)$\times$$10^5$~cm$^{-3}$, 
while the extended ($10^{\prime\prime}$$-$$17^{\prime\prime}$) component traces a warm ($T$$\sim$300$-$500~K) 
and dense ($n_{\rm H_2}$$\sim$(3$-$5)$\times$$10^7$~cm$^{-3}$) gas.
\item From a crude comparison between H$_2$O and CO (16$-$15) emission observed at R2, 
we estimate the H$_2$O/H$_2$ abundance of the warm and hot components to be 
(7$-$10)$\times$10$^{-7}$ and (3$-$7)$\times$10$^{-5}$.
\end{enumerate}
Our H$_2$O emission maps allow us, for the first time, to spatially
resolve
these two temperature components that were previously observed with HIFI and PACS.
We suggest that the compact hot component is associated with the jet that impacts 
the surrounding material, while the warm, dense, and extended one originates 
from the compression of ambient gas by the propagating flow.

\begin{acknowledgements}
\emph{Herschel} activities at INAF are financially supported by the ASI project 01/005/11/0.
HIFI has been designed and built by a consortium of 
institutes and university departments from across Europe, Canada and the 
United States under the leadership of SRON Netherlands Institute for Space
Research, Groningen, The Netherlands and with major contributions from 
Germany, France and the US. Consortium members are: Canada: CSA, 
U.Waterloo; France: CESR, LAB, LERMA, IRAM; Germany: KOSMA, 
MPIfR, MPS; Ireland, NUI Maynooth; Italy: ASI, IFSI-INAF, Osservatorio 
Astrofisico di Arcetri- INAF; Netherlands: SRON, TUD; Poland: CAMK, CBK; 
Spain: Observatorio Astron{\'o}mico Nacional (IGN), Centro de Astrobiolog{\'i}a 
(CSIC-INTA). Sweden: Chalmers University of Technology - MC2, RSS $\&$ 
GARD; Onsala Space Observatory; Swedish National Space Board, Stockholm 
University - Stockholm Observatory; Switzerland: ETH Zurich, FHNW; USA: 
Caltech, JPL, NHSC.
\end{acknowledgements}


\Online

\begin{appendix} 
\section{Additional lines observed with HIFI}

\begin{figure*}
\centering
\includegraphics[width=0.9\textwidth]{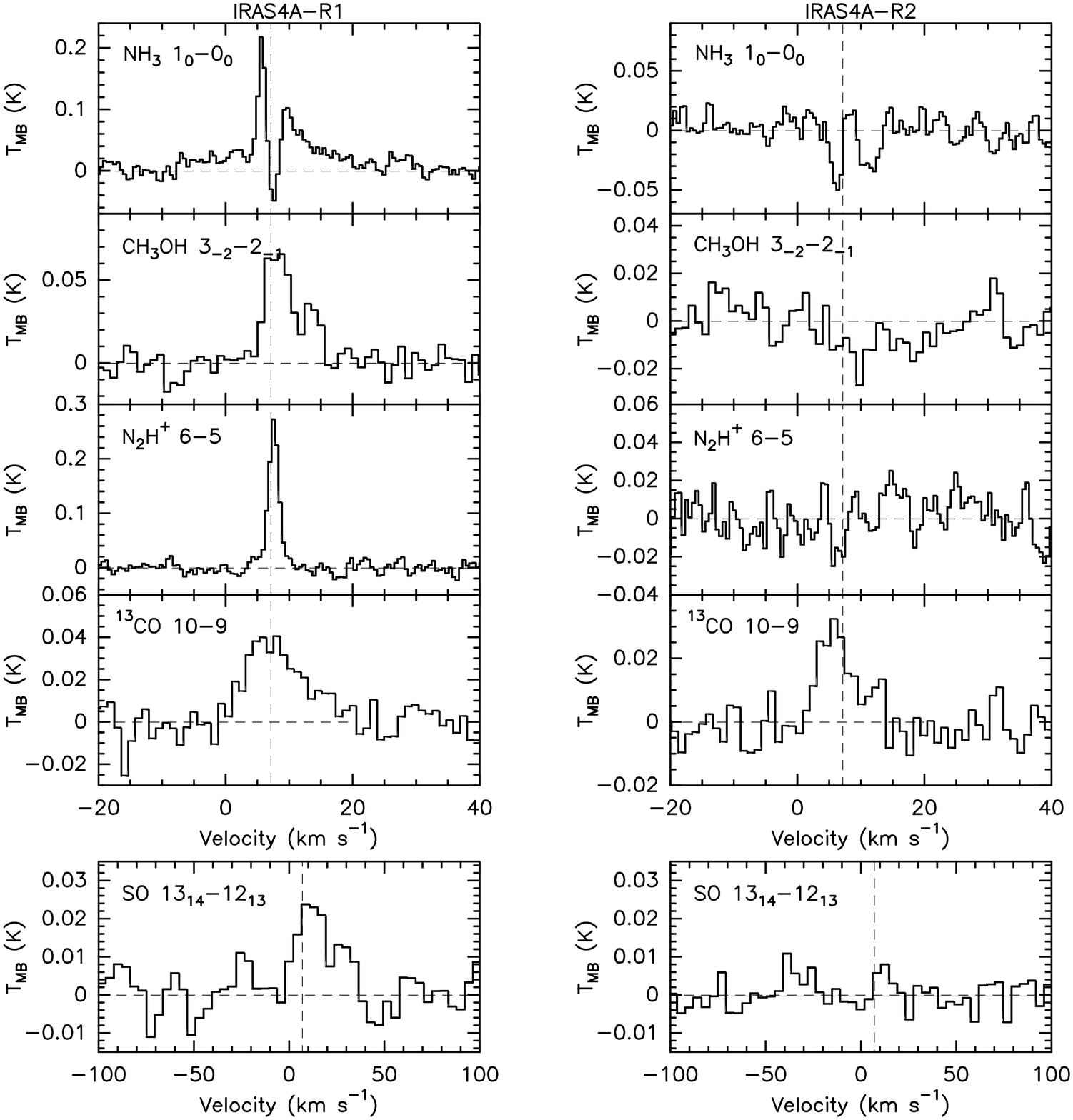}
\caption{Additional lines observed with HIFI at the selected shock positions.
All observations were taken in single-pointing mode, with the 
exception of the $^{13}$CO (10$-$9) line. The absorption features seen in the NH$_3$ spectrum at R2 
are due to contamination from emission in the off-source reference position.
}
\label{afig:OTHERlines}
\end{figure*}

The spectra of additional lines detected within the HIFI bands 
are presented in Fig.~\ref{afig:OTHERlines} and summarized in Table~\ref{table:HIFI_additional_lines}.
We note that the R1 shock position seems to be more chemically rich than R2; more molecules are detected at R1 than at R2.
The profiles of the lines detected at R1 show a red-wing component associated with the outflow 
and a low-velocity emission component possibly associated with the chemically enriched envelope material.
The latter is clearly associated with the detection of N$_2$H$^+$ and the 
absorption dip, centred at the source velocity, present in the NH$_3$ spectrum.
This detection is consistent with the beam size of the observations being large enough (about $38^{\prime\prime}$)
to encompass part of the emission associated with the central driving source of the outflow, 
along with the emission coming from the outflow itself.
We finally point out that the absorption features seen in the NH$_3$ spectrum at R2 
are due to contamination from emission in the off-source reference position.

\begin{table*}
\caption{Parameters of additional lines observed with HIFI.}
\label{table:HIFI_additional_lines}
\centering
\begin{threeparttable}
\renewcommand{\footnoterule}{}
\begin{tabular}{l c c r c c r r c c c } 
\hline\hline
Line & Observation ID & Band & Mode\tnote{a} & Sideband & $\eta_{\rm MB}$ & Frequency & $E_{\rm u}/k_{\rm B}$ & HPBW & R1 & R2  \\
&&&& && (MHz) & (K) & (arcsec) & \multicolumn{2}{c}{(x=yes)} \\
\hline
N$_2$H$^+$ (6$-$5)       & 1342248895, 1342248897 & 1 & SP & LSB & 0.75 &  558966.50   &    94 & 38  & x & --  \\
SO ($13_{14}$$-$$12_{13}$)     & 1342248895, 1342248897 & 1 & SP & LSB & 0.75 &  560178.65   & 193 & 38  & x & --  \\
CH$_3$OH ($3_{-2}$$-$$2_{-1}$) & 1342248895, 1342248897 & 1 & SP & USB & 0.75 &  568566.05   &  40 & 38  & x & --  \\ 
NH$_3$ ($1_0$$-$$0_0$)      & 1342248895, 1342248897 & 1 & SP & USB & 0.75 &  572498.16   &   27 & 37  & x & -- \\
$^{13}$CO (10$-$9)      & 1342249853, 1342249854  & 4 &  M & LSB & 0.74 & 1101349.66   &  291 & 19  & x & x  \\
\hline
\end{tabular}
\begin{tablenotes}
\item[a] SP = Single Pointing mode; M = Mapping mode. 
\end{tablenotes}
\end{threeparttable}
\end{table*}

\end{appendix}


\begin{thebibliography}{}

  \bibitem[Arce et al.(2007)]{arce2007} Arce, H.~G., Shepherd, D., Gueth, F., et al.\ 2007, Protostars and Planets V, 245 

  \bibitem[Baek et al.(2009)]{baek2009} Baek, C.~H., Kim, J., \& Choi, M.\ 2009, \apj, 690, 944 

  \bibitem[Bally et al.(1996)]{bally1996} Bally, J., Devine, D., \& Reipurth, B.\ 1996, \apjl, 473, L49 

  \bibitem[Bjerkeli et al.(2012)]{bjerkeli2012} Bjerkeli, P., Liseau, R., Larsson, B., et al.\ 2012, \aap, 546, A29 

  \bibitem[Blake et al.(1995)]{blake1995} Blake, G.~A., Sandell, G., van Dishoeck, E.~F., et al.\ 1995, \apj, 441, 689 

  \bibitem[Busquet et al.(2014)]{busquet2014} Busquet, G., Lefloch, B., Benedettini, M., et al.\ 2014, \aap, 561, A120 

  \bibitem[Ceccarelli et al.(2010)]{ceccarelli2010} Ceccarelli, C., Bacmann, A., Boogert, A., et al.\ 2010, \aap, 521, L22 

  \bibitem[Choi(2001)]{choi2001} Choi, M.\ 2001, \apj, 553, 219 

  \bibitem[Choi(2005)]{choi2005} Choi, M.\ 2005, \apj, 630, 976 
  
  \bibitem[Choi et al.(2011)]{choi2011} Choi, M., Kang, M., Tatematsu, K., Lee, J.-E., \& Park, G.\ 2011, \pasj, 63, 1281 

  \bibitem[Daniel et al.(2010)]{daniel2010} Daniel, F., Dubernet, M.-L., Pacaud, F., \& Grosjean, A.\ 2010, \aap, 517, A13 

  \bibitem[Daniel et al.(2011)]{daniel2011} Daniel, F., Dubernet, M.-L., \& Grosjean, A.\ 2011, \aap, 536, A76 

  \bibitem[de Graauw et al.(2010)]{degraauw2010} de Graauw, T., et al.\ 2010, \aap, 518, L6 
  
  \bibitem[Di Francesco et al.(2001)]{difrancesco2001} Di Francesco, J., Myers, P.~C., Wilner, D.~J., Ohashi, N., \& Mardones, D.\ 2001, \apj, 562, 770 

  \bibitem[Dionatos et al.(2013)]{dionatos2013} Dionatos, O., J{\o}rgensen, J.~K., Green, J.~D., et al.\ 2013, \aap, 558, A88 

  \bibitem[Dubernet et al.(2006)]{dubernet2006} Dubernet, M.-L., Daniel, F., Grosjean, A., et al.\ 2006, \aap, 460, 323 

  \bibitem[Dubernet et al.(2009)]{dubernet2009} Dubernet, M.-L., Daniel, F., Grosjean, A., \& Lin, C.~Y.\ 2009, \aap, 497, 911 

  \bibitem[Emprechtinger et al.(2013)]{emprechtinger2013} Emprechtinger, M., Lis, D.~C., Rolffs, R., et al.\ 2013, \apj, 765, 61 

  \bibitem[Flower \& Pineau des For{\^e}ts(2010)]{flower2010} Flower, D.~R., \& Pineau des For{\^e}ts, G.\ 2010, \mnras, 406, 1745 

  \bibitem[Giannini et al.(2001)]{giannini2001} Giannini, T., Nisini, B., \& Lorenzetti, D.\ 2001, \apj, 555, 40 

  \bibitem[Girart et al.(1999)]{girart1999} Girart, J.~M., Crutcher, R.~M., \& Rao, R.\ 1999, \apjl, 525, L109 

  \bibitem[Goicoechea et al.(2012)]{goicoechea2012} Goicoechea, J.~R., Cernicharo, J., Karska, A., et al.\ 2012, \aap, 548, A77 

  \bibitem[Green et al.(2013)]{green2013} Green, J.~D., Evans, N.~J., II, J{\o}rgensen, J.~K., et al.\ 2013, \apj, 770, 123 

  \bibitem[Herczeg et al.(2012)]{herczeg2012} Herczeg, G.~J., Karska, A., Bruderer, S., et al.\ 2012, \aap, 540, A84 

  \bibitem[Hirota et al.(2008)]{hirota2008} Hirota, T., Bushimata, T., Choi, Y.~K., et al.\ 2008, \pasj, 60, 37 

  \bibitem[Hollenbach \& McKee(1989)]{hollenbach1989} Hollenbach, D., \& McKee, C.~F.\ 1989, \apj, 342, 306 

  \bibitem[J{\o}rgensen et al.(2005)]{jorgensen2005} J{\o}rgensen, J.~K., Sch{\"o}ier, F.~L., \& van Dishoeck, E.~F.\ 2005, \aap, 437, 501 

  \bibitem[J{\o}rgensen et al.(2007)]{jorgensen2007} J{\o}rgensen, J.~K., Bourke, T.~L., Myers, P.~C., et al.\ 2007, \apj, 659, 479 

  \bibitem[Karska et al.(2013)]{karska2013} Karska, A., Herczeg, G.~J., van Dishoeck, E.~F., et al.\ 2013, \aap, 552, A141 

  \bibitem[Kaufman \& Neufeld(1996)]{kaufman1996} Kaufman, M.~J., \& Neufeld, D.~A.\ 1996, \apj, 456, 611 

  \bibitem[Knee \& Sandell(2000)]{kneesandell2000} Knee, L.~B.~G., \& Sandell, G.\ 2000, \aap, 361, 671 

  \bibitem[Kristensen et al.(2010)]{kristensen2010} Kristensen, L.~E., Visser, R., van Dishoeck, E.~F., et al.\ 2010, \aap, 521, L30 

  \bibitem[Kristensen et al.(2012)]{kristensen2012} Kristensen, L.~E., van Dishoeck, E.~F., Bergin, E.~A., et al.\ 2012, \aap, 542, A8 

  \bibitem[Kristensen et al.(2013)]{kristensen2013} Kristensen, L.~E., van Dishoeck, E.~F., Benz, A.~O., et al.\ 2013, \aap, 557, A23 

  \bibitem[Lay et al.(1995)]{lay1995} Lay, O.~P., Carlstrom, J.~E., \& Hills, R.~E.\ 1995, \apjl, 452, L73 

  \bibitem[Liseau et al.(1988)]{liseau1988} Liseau, R., Sandell, G., \& Knee, L.~B.~G.\ 1988, \aap, 192, 153 

  \bibitem[Looney et al.(2000)]{looney2000} Looney, L.~W., Mundy, L.~G., \& Welch, W.~J.\ 2000, \apj, 529, 477 

  \bibitem[Looney et al.(2007)]{looney2007} Looney, L.~W., Tobin, J.~J., \& Kwon, W.\ 2007, \apjl, 670, L131 

  \bibitem[Manoj et al.(2013)]{manoj2013} Manoj, P., Watson, D.~M., Neufeld, D.~A., et al.\ 2013, \apj, 763, 83 

  \bibitem[Maret et al.(2005)]{maret2005} Maret, S., Ceccarelli, C., Tielens, A.~G.~G.~M., et al.\ 2005, \aap, 442, 527 

  \bibitem[Maret et al.(2009)]{maret2009} Maret, S., Bergin, E.~A., Neufeld, D.~A., et al.\ 2009, \apj, 698, 1244 

  \bibitem[Nisini et al.(2010)]{nisini2010} Nisini, B., Benedettini, M., Codella, C., et al.\ 2010, \aap, 518, L120 

  \bibitem[Nisini et al.(2013)]{nisini2013} Nisini, B., Santangelo, G., Antoniucci, S., et al.\ 2013, \aap, 549, A16 

  \bibitem[Ott(2010)]{ott2010} Ott, S.\ 2010, Astronomical Data Analysis Software and Systems XIX, 434, 139 

  \bibitem[Pilbratt et al.(2010)]{pilbratt2010} Pilbratt, G.~L., Riedinger, J.~R., Passvogel, T., et al.\ 2010, \aap, 518, L1 

  \bibitem[Ray et al.(2007)]{ray2007} Ray, T., Dougados, C., Bacciotti, F., Eisl{\"o}ffel, J., \& Chrysostomou, A.\ 2007, Protostars and Planets V, 231 

  \bibitem[Roelfsema et al.(2012)]{roelfsema2012} Roelfsema, P.~R., Helmich, F.~P., Teyssier, D., et al.\ 2012, \aap, 537, A17 

  \bibitem[Sandell et al.(1991)]{sandell1991} Sandell, G., Aspin, C., Duncan, W.~D., Russell, A.~P.~G., \& Robson, E.~I.\ 1991, \apjl, 376, L17 

  \bibitem[Santangelo et al.(2012)]{santangelo2012} Santangelo, G., Nisini, B., Giannini, T., et al.\ 2012, \aap, 538, A45 

  \bibitem[Santangelo et al.(2013)]{santangelo2013} Santangelo, G., Nisini, B., Antoniucci, S., et al.\ 2013, \aap, 557, A22 

  \bibitem[Sch{\"o}ier et al.(2005)]{schoier2005} Sch{\"o}ier, F.~L., van der Tak, F.~F.~S., van Dishoeck, E.~F., \& Black, J.~H.\ 2005, \aap, 432, 369 

  \bibitem[Suutarinen et al.(2014)]{suutarinen2014} Suutarinen, A.~N., Kristensen, L.~E., Mottram, J.~C., Fraser, H.~J., \& van Dishoeck, E.~F.\ 2014, \mnras, 440, 1844 

  \bibitem[Tafalla et al.(2013)]{tafalla2013} Tafalla, M., Liseau, R., Nisini, B., et al.\ 2013, \aap, 551, A116 

  \bibitem[van der Tak et al.(2007)]{vandertak2007} van der Tak, F.~F.~S., Black, J.~H., Sch{\"o}ier, F.~L., Jansen, D.~J., \& van Dishoeck, E.~F.\ 2007, \aap, 468, 627 

  \bibitem[van Dishoeck et al.(2011)]{vandishoeck2011} van Dishoeck, E.~F., Kristensen, L.~E., Benz, A.~O., et al.\ 2011, \pasp, 123, 138 

  \bibitem[Vasta et al.(2012)]{vasta2012} Vasta, M., Codella, C., Lorenzani, A., et al.\ 2012, \aap, 537, A98 

  \bibitem[Y{\i}ld{\i}z et al.(2012)]{yildiz2012} Y{\i}ld{\i}z, U.~A., Kristensen, L.~E., van Dishoeck, E.~F., et al.\ 2012, \aap, 542, A86 

  \bibitem[Y{\i}ld{\i}z et al.(2013)]{yildiz2013} Y{\i}ld{\i}z, U.~A., Kristensen, L.~E., van Dishoeck, E.~F., et al.\ 2013, \aap, 556, A89 

\end{thebibliography}
\end{document}